\documentclass[sigconf]{acmart}
\usepackage{enumitem}
\usepackage{subfigure}
\usepackage{graphicx}
\usepackage{epsfig}
\usepackage{color}
\usepackage[ruled,vlined]{algorithm2e}

\newcommand{\eat}[1]{}
\newcommand{\inp}{\vskip .5em}

\begin{document}

\copyrightyear{2019}
\acmYear{2019}
\setcopyright{acmlicensed}
\acmConference{Conference}{}{Date, Year, Location.}
\acmPrice{15.00}
\acmDOI{http://dx.doi.org/10.1145/XXXXXXX.YYYYYYY}
\acmISBN{ISBN ISBN}

\fancyhead{}
\settopmatter{printacmref=false, printfolios=false}

%IF NEED BE.
% --- End of Author Metadata ---

\title{Meet Cyrus -- The Query by Voice Mobile Assistant for the Tutoring and Formative Assessment of SQL Learners}
\titlenote{Research supported in part by a STEM Center grant, and a National Science Foundation grant DRL 1515550.}
%\subtitle{Extended Abstract}
%\subtitlenote{The full version of the author's guide is available as
%  \texttt{acmart.pdf} document}

\author{Josue Espinosa Godinez}
\affiliation{%
  \institution{Department of Computer Science\\ University of Idaho, USA}
%  \streetaddress{University of Idaho, USA}
%  \city{Moscow}
%  \state{Idaho}
%  \postcode{83844, USA}
}
\email{espi9890@vandals.uidaho.edu}

\author{Hasan M. Jamil}
\orcid{0000-0002-3124-3780}
\affiliation{%
 \institution{Department of Computer Science\\ University of Idaho, USA}
%  \streetaddress{University of Idaho, USA}
%  \city{Moscow}
%  \state{Idaho}
%  \postcode{83844, USA}
}
\email{jamil@uidaho.edu}

% The default list of authors is too long for headers}
\renewcommand{\shortauthors}{Hasan Jamil}
\renewcommand{\shorttitle}{Cyrus -- The Query by Voice Assistant}

\begin{abstract}
Being declarative, SQL stands a better chance at being the programming language for conceptual computing next to natural language programming. We examine the possibility of using SQL as a back-end for natural language database programming. Distinctly from keyword based SQL querying, keyword dependence and SQL's table structure constraints are significantly less pronounced in our approach. We present a mobile device voice query interface, called {\em Cyrus}, to arbitrary relational databases. Cyrus supports a large type of query classes, sufficient for an entry level database class. Cyrus is also application independent, allows test database adaptation, and not limited to specific sets of keywords or natural language sentence structures. It's cooperative error reporting is more intuitive, and iOS based mobile platform is also more accessible compared to most contemporary mobile and voice enabled systems.
\end{abstract}

\ccsdesc[500]{Human-centered computing~Natural language interfaces}
\ccsdesc[300]{Human-centered computing~Graphical user interfaces}
\ccsdesc[300]{Human-centered computing~Auditory feedback}
\ccsdesc[100]{Human-centered computing~HCI theory, concepts and models}
\ccsdesc[100]{Human-centered computing~Empirical studies in HCI}
\ccsdesc[500]{Computing methodologies~Information extraction}
\ccsdesc[300]{Computing methodologies~Speech recognition}
\ccsdesc[300]{Computing methodologies~Knowledge representation and reasoning}
\ccsdesc[300]{Theory of computation~Database query languages (principles)}

\keywords{Query by Voice;
SQL tutoring;
query mapping;
formative assessment;
self-paced learning;
mobile learning system}

\maketitle

\section{Introduction}
\label{introduction}

Although declarative programming is arguably more intuitive, students still find the transition from imperative languages to SQL hugely difficult. They often struggle to form complex queries that are rich in semantic nuances, especially those involving nested sub-queries or {\sf GROUP BY} functions \cite{AhadiPBL15s}. To help students learn SQL better, researchers have been trying to develop various teaching systems and learning strategies, and support them with powerful online \cite{Prior14} and desktop tools \cite{BrusilovskySYLZZ10}. Despite the recognition of its importance, research in developing teaching tools for SQL, especially smart tools have been scant \cite{MitrovicO16,Prior14}. Most of the contemporary systems focus on syntactic aspects and do not go far in teaching the semantic and conceptual underpinnings of declarative programming using SQL with the few exceptions of the automated SQL exercise grading \cite{KleinerTH13s} and guided teaching \cite{LavbicMZ17} systems.

Recent introduction of voice services such as Amazon {\em Alexa}, Apple {\em Siri} and {\em Google Assistant} leveraged extensive research in understanding and mapping natural language queries (NLQ) to SQL \cite{LiJ16} to substantially simplify access to data and to help users who otherwise would not use the vast amount of knowledge to improve their lives \cite{ReisPPB17s}. Internet of things (IoT) and smart devices research are also encouraging people to use various digital home services and mobile phone applications using voice recognition and speech processing for automated information gathering from databases. These successful voice technologies remain largely unexplored in educational systems for teaching SQL to first-time learners. We believe that integrating voice or natural language based query capabilities in online systems will make learning platforms powerful. If such a system can be coupled with automated tutoring and grading functions, it can serve as a smart and comprehensive learning environment for SQL.

\subsection{Voice Assistant as a Teaching Aid}

In the recent years, there has been a significant shift in the size, technology, speed and cost of hardware devices for the better. Today, the ubiquity of smart and mobile devices with large memories offers us the opportunity to rethink how educational aids may be designed for the new century. It is estimated that there are about 700 million Apple iPhones in use worldwide and this number could reach to 1 billion in a few years. With the unprecedented accessibility afforded by such devices, it is important to consider their large-scale applications in education, in particular, since a large number of them are in the hands of students. It is thus natural to explore the opportunity to help students learn using mobile devices, and especially using effortless, convenient and ad hoc voice technology to support uninhibited exploration.

A substantial segment of users are also already familiar with voice-interfaces and with prominent virtual assistants such as Siri. It is thus conceivable that a voice enabled SQL tutoring and assessment tool along the lines of the systems such as {\em PhotoMAT} \cite{ShelleyD0LM15} or \cite{ChenC09} could support anytime online learning in a hands free manner, and with smart response read out options, students could self-assess their SQL composition abilities over known test databases. Naturally, platforms such as iOS with preexisting APIs for NLP, speech-to-text, and speech synthesis are ideal due to their first-party support, large user-base, and familiarity of voice-interface systems. With the combination of widespread device availability, inexpensive hardware cost (relative to specialized hardware), and stable and heavily field-tested firmware, it is a prudent choice to utilize this platform to implement such a system.

\subsection{Advantages of a Voice Assistant}

While the usefulness of voice interactions for interfacing with applications and databases are somewhat well understood, its use as a teaching aid requires some justifications mainly because not many research have used this technology for teaching systems. It is argued that voice interfacing to databases in general provides three main advantages \cite{LyonsTBCK16s} -- hands-free access, personalizable vocabulary and dialogue-based querying. Given a set of learning objectives of SQL query constructs and classes, all a student possibly wants is to see whether the system generated SQL expression for a query in English matches with his mental formulation of the query in SQL, and produces identical response. From this standpoint, hands-free access to database querying engines and personalizable vocabulary certainly could play major roles in a mobile SQL tutoring and assessment system, which are the major focus of this research.

\section{Related Research}

Although text interfaces to databases \cite{LiJ16} have been explored relatively more than voice interfaces \cite{KumarKMS13}, their use for teaching SQL is less explored \cite{Prior14}. While we can point to SQL teaching systems such as \cite{BrusilovskySYLZZ10,MitrovicO16,KleinerTH13s,LavbicMZ17,Prior14}, we are unable to discuss them in this paper for the want of space except {\em EchoQuery} \cite{LyonsTBCK16s} as this framework has much in common with our approach in Cyrus. In EchoQuery, users are able to communicate with the database using voice at any time and queries can be asked incrementally, in steps, in a context of prior queries and stateful dialogue-based refinement are supported along with clarification if queries are incomplete or ambiguous. Finally, EchoQuery allows for a personalized vocabulary on a per-user basis, which makes the system robust. Although Cyrus is not as adept at continuous querying, Cyrus does have a more vigorous translation system for matching user queries over heterogeneous schemes and handle numbers in NLQ better. Cyrus can also map column/table names that span multiple words with special delimiters such as ``Track ID'' to ``Track\_id'' or ``TrackId'' by maintaining a history stack with context.

\section{Cyrus User Interface}

\begin{figure}[ht]
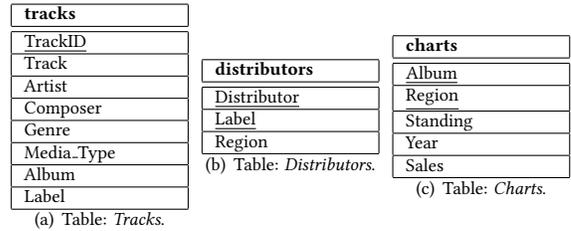

{\footnotesize
\subfigure[Table: {\em Tracks}.]
{\label{tt}
\begin{tabular}{|p{2cm}|}
\hline
{\bf tracks}\\ \hline\hline
\underline{TrackID}\\ \hline
Track\\ \hline
Artist\\ \hline
Composer\\ \hline
Genre\\ \hline
Media\_Type\\ \hline
Album\\ \hline
Label\\ \hline
\end{tabular}
}
\subfigure[Table: {\em Distributors}.]
{\label{dt}
\begin{tabular}{|p{2cm}|}
\hline
{\bf distributors}\\ \hline\hline
\underline{Distributor}\\ \hline
\underline{Label}\\ \hline
Region\\ \hline
\end{tabular}
}
\subfigure[Table: {\em Charts}.]
{\label{ct}
\begin{tabular}{|p{2cm}|}
\hline
{\bf charts}\\ \hline\hline
\underline{Album}\\ \hline
\underline{Region}\\ \hline
Standing\\ \hline
Year\\ \hline
Sales\\ \hline
\end{tabular}
}
\caption{Scheme of {\em Music} Database.} \label{DB-Scheme}}
\end{figure}

Cyrus is a smart tutoring and assessment system for SQL designed on the mobile iOS platform. The voice enabled interface translates natural English language queries on a test database into executable SQL queries. A successful translation shows the translated query and the computed results for the student to review and validate her mental model of the query. Since Cyrus generates editable SQL queries that can be executed optionally, students have the opportunity to master the language through iterative refinement. As a formative assessor or grader, Cyrus also compares system mapped queries with students' translated queries over a test database. In the role of a grader, to be deemed correct, a student SQL query must compute identical views as does the system generated query. The test queries are crafted by an instructor as database assignments with escalating difficulty levels. In the remainder of this paper, we use a music database as shown (with the primary keys underlined for each table) in figure  \ref{DB-Scheme} to exemplify the features of Cyrus and for other illustrative purposes. While we sketch how Cyrus maps NLQ to SQL in section \ref{mapnlq} and use examples to discuss the functionalities of Cyrus on intuitive grounds, we defer the full treatment of the NLQ to SQL mapping algorithm to an extended paper.

\subsection{Using Cyrus}

\begin{figure}[h]
	\centering
	\subfigure[Cyrus in tutoring mode.]{\label{fig:tutoring}\includegraphics[scale=0.309375]{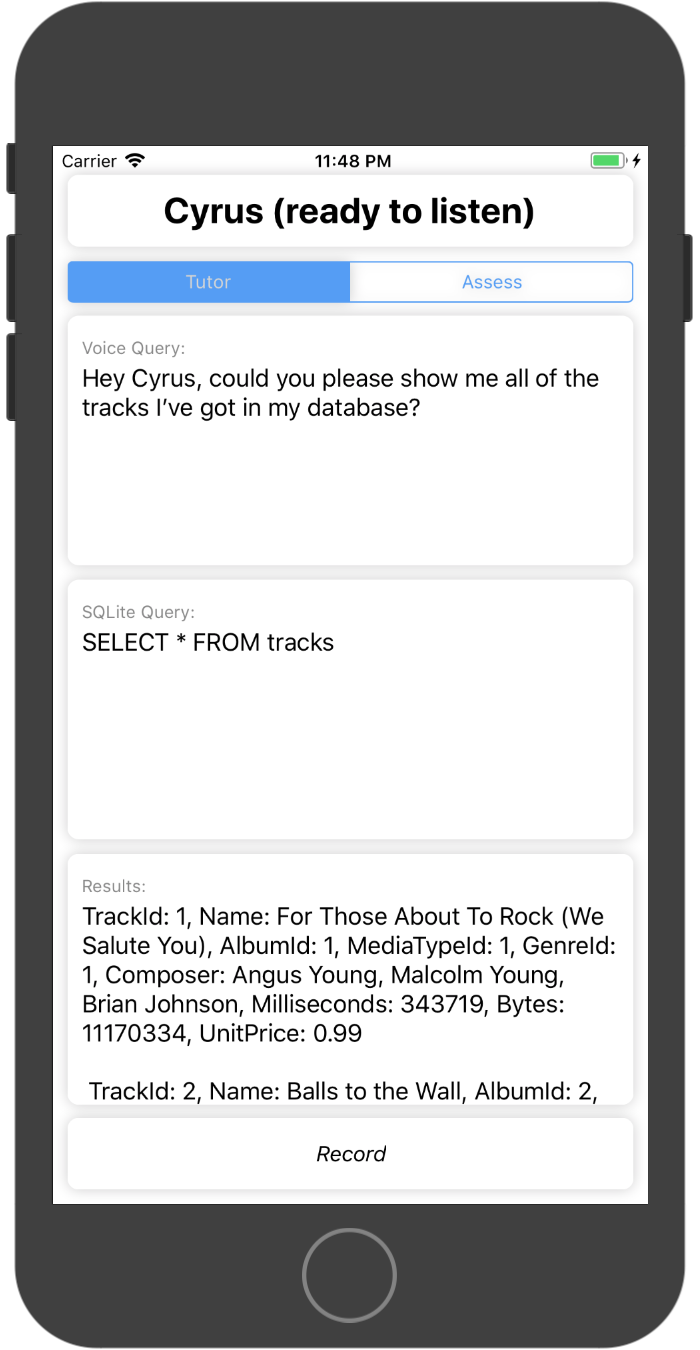}}
	\subfigure[Cyrus in assessment mode.]{\label{fig:assessment}\includegraphics[scale=0.309375]{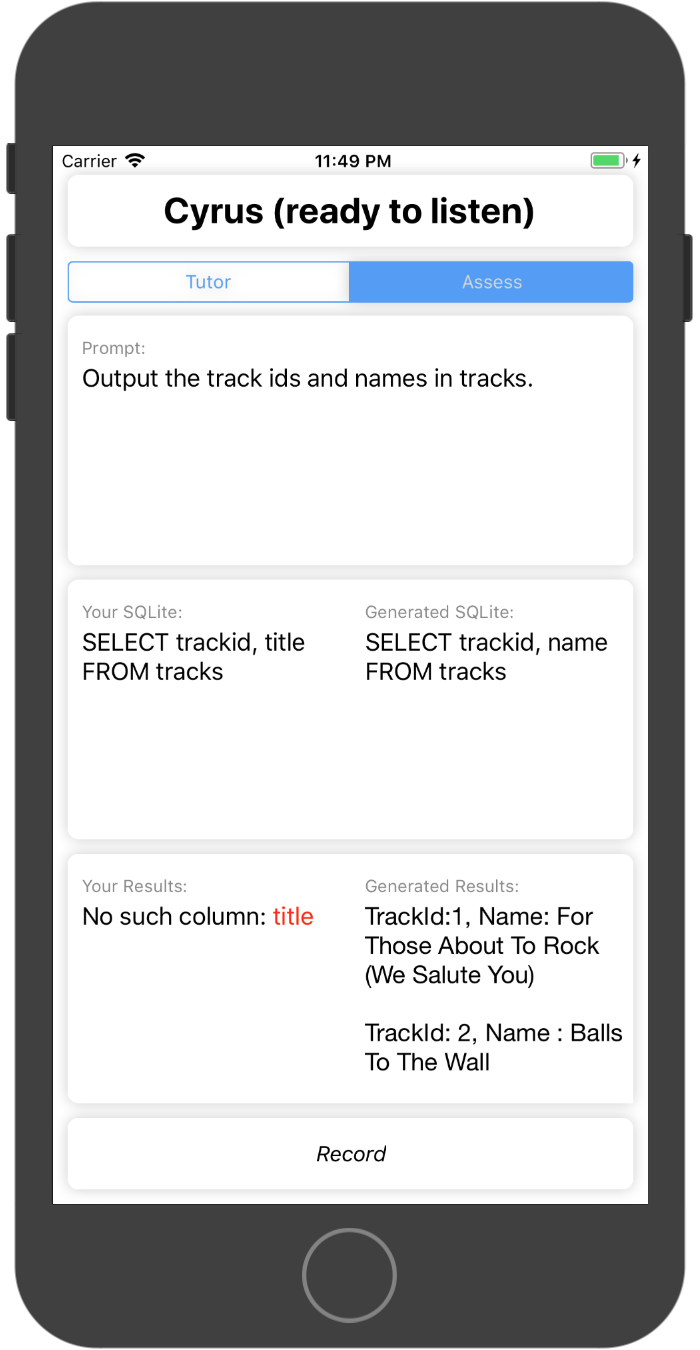}}
	\caption{Cyrus interface.}
	\label{fig:interface}
\end{figure}

Cyrus has two main modes, as a tutor, it allows students to choose an example database, and in assessment mode, it additionally allows to choose a difficulty level. In tutoring mode, it simply accepts voice query in natural English, maps the query to SQL, executes it, and shows the computed response and the SQL query it executed to produce the result. Students are allowed to edit the SQL query or write their own, bypassing the voice interface, for execution. In assessment mode though, it shows only the test queries in English at the difficulty level chosen by the student to transcribe in SQL. However, in assessment mode, the voice interface is disabled leaving only the text interface for the students to write the SQL queries without system assisted translation from NLQ. A score is shown at the end of the session informing the student how well she did according to the grading scheme of the instructor. Figure \ref{fig:interface} shows the Cyrus interface in both tutoring and assessment modes.

\subsection{Supported Query Classes}

In the current Cyrus prototype, our focus has been to support a sufficient number of SQL  query classes for an entry level database class, and allow multiple natural language renditions of the queries to support variability and personalization. Students with different levels of spoken English skills (e.g., especially students with non-English native languages) are expected to express differently in free form languages, although the corresponding SQL renditions are usually limited. Although in formative and summative assessment mode, written queries will be framed by the course instructors, in tutoring mode, the students are expected to frame their own queries just by looking at the database schema and the instance. Thus, the expressions of the queries in English will vary from student to student necessitating a more flexible natural language input support. In the sections below, we use the music database scheme of figure \ref{DB-Scheme} to illustrate the functionalities of Cyrus and discuss the supported query classes on intuitive grounds. We note that while Cyrus is able to map most queries, in the event it is presented with a query for which it cannot find a mapping, Cyrus will ask to restate the query differently in order to retry.

\subsubsection{Wildcard Queries}

Consider, for example, the query {\em ``List all music tracks''} over the music database. Obviously, the information for all music albums are in the {\em tracks} table, and a wildcard query will return the expected response. But, unlike other systems, this query can be asked in multiple different ways in Cyrus, including the following widely different forms, and all will still map to the wildcard SQL query below.
\inp
\indent {\sf SELECT *\\
\indent FROM} {\em tracks};
\inp
\begin{enumerate}
\item {\em ``Hey Cyrus, could you please show me all of the tracks I've got in my database?''}
\label{q0}
\item {\em ``What are the songs in the database?''}
\label{q1}
\item {\em ``List all of my tracks for me.''}
\item {\em ``Tracks, please.''}
\end{enumerate}
\noindent In Cyrus, indeed an exceedingly large number of verbal syntax are allowed so long sufficient keywords are used that can be successfully related to a table name in the chosen database. In particular, query \ref{q1} above did not use the keyword {\em tracks}, yet Cyrus is able to relate the {\em tracks} table to the songs keyword as a personalization feature leveraging a synonym/vocabulary database.

\subsubsection{Projection Queries}

Often, specific field of tuples or columns are of interest and need to be extracted. Cyrus supports such extractions as projection queries. However,
this innocuous extension of wildcard queries is relatively harder to map and its simplicity is largely deceptive. Consider the following natural language queries. These queries map to the SQL query
\inp
\indent {\sf SELECT} {\em TrackId, Track} \\
\indent {\sf FROM} {\em tracks};
\inp
\begin{enumerate}[resume]
\item {\em ``Show track id and name from the tracks table.''}
\item {\em ``Output the track ids and names in tracks.''}
\item {\em ``List the number and title of the songs.''}
\label{q2}
\item {\em ``Track id, name, tracks.''}
\end{enumerate}
In these queries, Cyrus primarily looks for a table name to match followed by matching column names, uniquely. Note that, the synonyms and similarity functions it uses makes the matching process complicated. Conceptually, it uses a maximizing function based on Levenshtein Distance (discussed in section \ref{distance}) to find a table and attribute list pair that uniquely matches the terms in the English query. In particular, query \ref{q2} above still maps to the SQL query above because {\em number} matches with {\em TrackId}, {\em title} with {\em Track} and {\em songs} with {\em tracks}, and the combined similarity of the pair $\langle${\em songs}, \{{\em number, title\}}$\rangle$ is much greater for the pair $\langle${\em tracks}, \{{\em TrackId, Track\}}$\rangle$ than any other pairs.

\subsubsection{Selection Queries}
\label{selq}

Queries satisfying constraints are also supported in Cyrus. Consider the SQL query
\inp
\indent {\sf SELECT * \\
\indent FROM} {\em tracks} \\
\indent {\sf WHERE} {\em TrackId=1479} {\sf AND} {\em Composer='Jimi Hendrix'};
\inp
\noindent which lists the track details for the song with the {\em TrackID} 1479 for composer {\em Jimi Hendrix}, in the table {\em tracks}. This query, as usual, can be asked in any of the following forms, among many other.
\begin{enumerate}[resume]
\item {\em ``Get tracks composer Jimi Hendrix where the track id is 1479.''}
\item {\em ``Print Jimi Hendrix composed songs with the number 1479.''}
\label{q3}
\item {\em ``Show me the tracks composed by Jimi Hendrix if the track id is 1479.''}
\label{q5}
\item {\em ``Tracks composer Jimi Hendrix 1479 track id.''}
\end{enumerate}
Cyrus maps these voice commands to the selection query above using techniques similar to the projection queries. Projection of course can be easily combined with a selection query, e.g., the SQL query below is a mapping from the NLQ that follows.
\inp
\indent {\sf SELECT} {\em Track, Media\_Type, Genre} \\
\indent {\sf FROM} {\em tracks} \\
\indent {\sf WHERE} {\em TrackId=1479} {\sf AND} {\em Composer='Jimi Hendrix'};
\begin{enumerate}[resume]
\item {\em ``Print track name, media type and genre of Jimi Hendrix composed songs whenever the number is 1479.''}
\label{q4}
\end{enumerate}

\subsubsection{Join or Multi-Relational Queries}
Join queries, often called multi-relational or SPJ (select-project-join) queries, are in fact the most general and common type of query classes in relational databases. Such queries require linking more than one relations to form a large table on which the selection conditions and projections are applied to find responses. For example, consider the queries
\begin{enumerate}[resume]
\item {\em ``List all the artists and their albums distributed by Redeye Distribution in USA that charted top 5 in USA in 2017.''}
\item {\em ``List top 5 ranked 2017 albums and artists distributed by Redeye Distribution in USA.''}
\end{enumerate}
Cyrus constructs the join query below over the scheme in figure \ref{DB-Scheme} for either of the two NLQs above.
\inp
\indent {\sf SELECT} {\em Album, Artist} \\
\indent {\sf  FROM} {\em tracks} {\sf NATURAL JOIN} {\em distributors} {\sf NATURAL JOIN} {\em charts} \\
\indent {\sf WHERE} {\em Distributor='Redeye Distribution'} {\sf AND} {\em Year=2017} \\ \indent \hspace*{2mm} {\sf AND} {\em Standing $\leq$ 5};
\inp
\noindent From our example database, we compute the response as {\em Reflection} by {\em Brian Eno} and {\em Take Me Apart} by {\em Kelela} which ranked higher than 6 in multiple charts in 2017 under the label {\em Warp}, which can be computed only after joining these three tables. Although {\em Redeye} distributed other albums such as {\em Death Peak} by {\em Clark, Shake the Shudder} by {\em !!!}, and {\em London 03.06.17} by {\em Aphex Twin}, they did not make the top 5 chart anywhere in 2017.

\subsubsection{Division Queries} Consider the queries below.
\begin{enumerate}[resume]
\item {\em ``List artists who recorded albums under all the labels artist Gone is Gone has ever recorded.''}
\label{q6}
\item {\em ``List all the artists who have recorded albums at least with all the labels who recorded Gone is Gone too.''}
\end{enumerate}
In such queries, we are interested to find association of one object with a set of objects that cannot be computed using one simple join query -- called the {\em division} queries. While it can be computed in several steps without using the concept of division, Cyrus maps these queries to the nested SQL query below utilizing SQL's tuple variable feature. Recognizing and mapping such a query is one of the most difficult ones in NLQ to SQL translation.
\inp
\indent {\sf SELECT} {\em Artist} \\
\indent {\sf  FROM} {\em tracks} {\sf AS} {\em t}\\
\indent {\sf WHERE} \begin{tabular}{l}
({\sf SELECT} {\em Label}\\
{\sf  FROM} {\em tracks} \\
{\sf WHERE} {\em Artist=t})
\end{tabular} {\sf  CONTAINS} \begin{tabular}{l}
({\sf SELECT} {\em Label}\\
{\sf  FROM} {\em tracks} \\
{\sf WHERE} {\em Artist=}\\ \indent {\em 'Gone is Gone'})
\end{tabular}

\subsubsection{Aggregate Queries with Sub-Group Filtering}
In contrast, aggregate queries insist on creating groups on which aggregate operations such as {\bf sum}, {\bf avg}, {\bf max}, {\bf min}, etc. are computed and filter conditions applied. For example, consider the query
\begin{enumerate}[resume]
\item {\em ``Print the artists who sold more than 2 million copies of their albums in USA in 2017.''}
\label{q7}
\end{enumerate}
This query requires filtering out first only those albums sold in 2017 followed by creating a subgroup ({\sf GROUP BY}) of albums with their sales record to find the total sales and then only selecting those artists who has more than 2 million in sales within this filtered group ({\sf HAVING} clause). On our example database, Cyrus translates this NLQ to the following SQL query.
\inp
\indent {\sf SELECT} {\em Artist}, {\sf SUM}({\em Sales}) \\
\indent {\sf  FROM} {\em tracks} {\sf NATURAL JOIN} {\em charts}\\
\indent {\sf WHERE} {\em Year=2017} \\
\indent {\sf  GROUP BY} {\em Artist} \\
\indent {\sf HAVING} {\sf SUM}({\em Sales}) $>$ {\em 2,000,000}

\section{Cyrus System Architecture and Implementation}
\label{algorithms}

Voice processing apps are popular on mobile platforms and mobile app operating systems such as Apple iOS, Microsoft Phone OS, HP WebOS and Google Android are supporting an increasing number of such tools for app developers. These tool support includes voice-to-text, speech analysis, and text-to-speech libraries. One of the leading intelligent voice processing system is Apple's Siri on iOS platform. In iOS in particular, Apple supports {\em NSLinguisticTagger}, {\em AVSpeechSynthesizer}, and {\em SFSpeechRecognizer} classes in its Foundation and Speech frameworks suit that are relatively more mature and stable. {\em NSLinguisticTagger} is a uniform interface that can be used for a variety of natural language processing functions such as segmenting natural language text into paragraphs, sentences, or words, and tag information about those segments, such as part of speech, lexical class, lemma, script, and language. The {\em AVSpeechSynthesizer} class, on the other hand, can be used to produce synthesized speech from text, while the {\em SFSpeechRecognizer} class help recognize a speech of a locale. Availability of these enabling tools was the primary motivation for choosing iOS platform for Cyrus along with the fact that Apple iPhone is the most widely used mobile smart device.

At a high level, the Cyrus query processing pipeline proceeds in five distinct steps - voice (query) acquisition, voice to text transcription, text parsing, text to SQL mapping, and SQL query processing. In Cyrus, we have assembled the pipeline leveraging iOS speech library classes embedded within our query processing system. We describe the process below in some detail but defer a full discussion to an extended version of this article.

\subsection{Voice Query Acquisition and Transcription} In iOS platform, {\em SFSpeechRecognizer} requires the use of permission through its {\em requestAuthorization} function for app permission to perform speech recognition, followed by a monitoring session using {\em SFSpeechRecognitionTask} via the {\em AVAudioSession} to create a new live recognition request using {\em SFSpeechAudioBufferRecognitionRequest}. This recognition task then keeps track of the user's speech and utilize a result handler to generate the most accurate text transcription possible until the recognition request has been completed. While these functions instantiate to local language by default, we have used English initialization to limit variability.

\subsection{Parsing Natural Language Queries}

Once the text transcription is received, semantic understanding of the text query can begin using natural language processing. We preprocess the text using a process called {\em lemmatization} that standardizes the text query. For this purpose, we use
the powerful {\em NSLinguisticTagger} class which is able to perform language identification, tokenization, lemmatization, part of speech (PoS) tagging, and named entity recognition tasks with proper instantiations. As mentioned earlier, we instantiate our language to English and proceed with the remaining steps in parsing the text. After tokenization, a lemmatization step is performed to reduce inflectional forms and often derivationally related forms of words to a common base form. For example, words such as {\em am, are,} and {\em is} are replaced with {\em be}, a stem form of a word token, to help transcribe sentences such as {\em  the boy's cars are different colors} to {\em the boy car be differ color}. Lemmatization help in situations when inflectional and derivational forms force us to search for too many words and establish their semantic meanings making understanding difficult although the words involved differ in their flavor.

Before we initiate PoS tagger in {\em NSLinguisticTagger} to obtain a tagged string $Q_p$, we include tagger option {\em .joinNames} to collapse ``San Francisco'' to ``SanFrancisco'' to be able to treat this token as a singular entity instead of two separate words, and enumerate the tags within the string using a name type and lexical class scheme, to aid semantic matching of the tokens with the database scheme in our next step.

\subsection{Text Processing for Schema Recognition}
\label{mapnlq}

Mapping the PoS tagged and enriched text query is a complex process. In Cyrus, we follow a heuristic approach for the identification of table names for the {\sf FROM} clause, attribute lists in the {\sf SELECT} clause and Boolean conditions in the {\sf WHERE} clause of all SQL queries. This approach also helps us avoid complex grammatical and semantic analysis of the text query that may not bear fruit at the end anyway. But by doing so we risk failure even on a otherwise mappable semantically correct query.

\subsubsection{Levenshtein Distance for String Matching}

To match entities referenced in the English query with the table names in the database, and to map possible properties of entities to attribute names of tables, we leverage the popular Levenshtein distance for term matching, which basically is a edit distance function with some interesting properties, including triangle inequality. Mathematically, Levenshtein distance between two strings $a$ and $b$ is a function of the form
\begin{equation*}
lev_{a,b}(i,j) =
\begin{cases}
\max(i,j) & \text{if } \min(i,j)=0,\\
\min \begin{cases}
     lev_{a,b}(i-1,j)+1\\
     lev_{a,b}(i,j-1)+1\\
     lev_{a,b}(i-1,j-1) + \\ \hspace{20mm} 1_{(a_i\not = b_j)}
     \end{cases} & \text{otherwise.}
\end{cases}
\end{equation*}
where  $ i=|a|$ and $j=|b|$ are lengths of strings $a$ and $b$, and $1_{(a_i\not = b_j)}$ is the indicator function which equals to 0 when $a_i = b_j$ and equal to 1 otherwise, and $lev_{a,b}(i,j) $ is the distance between the first $i$ characters of $a$ and the first $j$ characters of $b$. Note that the first element in the minimum corresponds to deletion (from $a$ to $b$), the second to insertion and the third to match or mismatch, depending on whether the respective symbols are the same. As an example, consider the strings ``eaten'' and ``sitting.'' We can transform or edit ``eaten'' in five steps,
\begin{quote}
eaten $\stackrel{1}{\rightarrow}$ saten $\stackrel{2}{\rightarrow}$ siten $\stackrel{3}{\rightarrow}$ sittn $\stackrel{4}{\rightarrow}$ sittin $\stackrel{5}{\rightarrow}$ sitting by first substituting ``s'' for ``a'', then substituting ``i'' for ``a'', then substituting ``t'' for ``e'', then inserting ``i'' before ``n", and finally inserting ``g'' at the end.
\end{quote}
to make it ``sitting,'' i.e., $lev_{eaten,sitting}(|eaten|,|sitting|)=5$. We use a special case of $lev_{a,b}(i,j) =0$ when the strings $a$ and $b$ are identical or one is a substring of the other.

\subsubsection{Table and Column Name Recognition}
\label{distance}

\begin{figure}[h]
	\centering
	\includegraphics[scale=0.425]{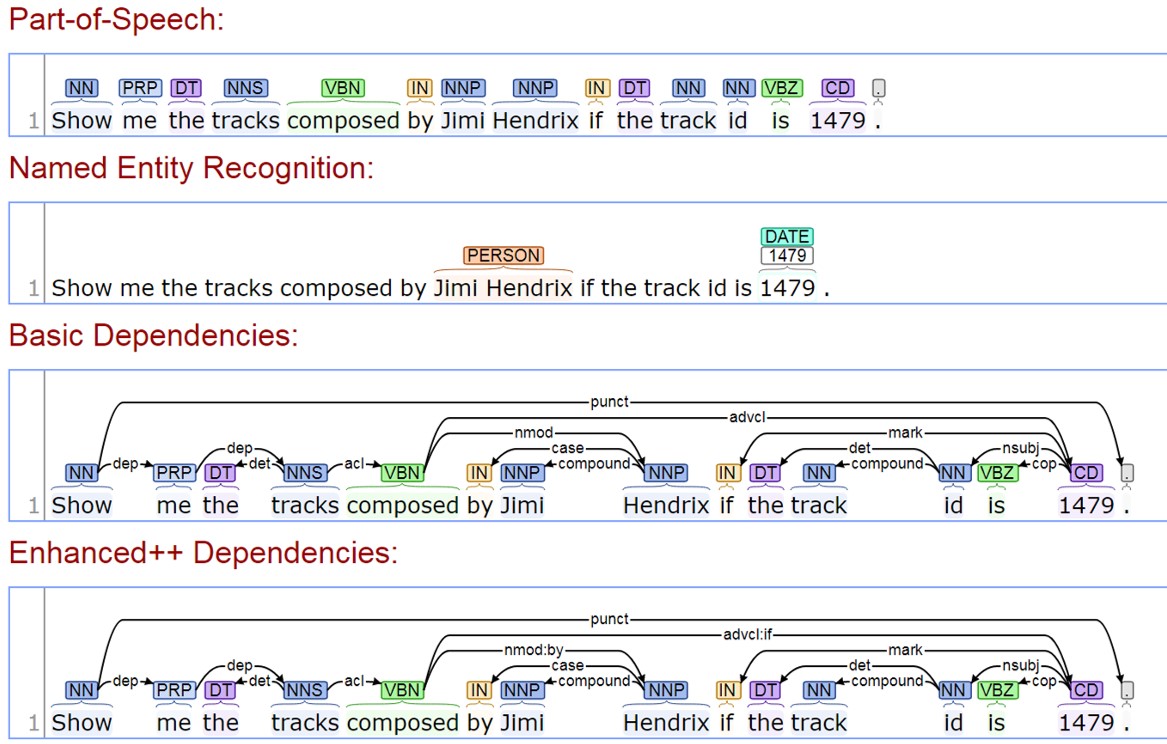}
	\caption{Parsing query \ref{q5} using NSLinguisticTagger.}
	\label{PosTag}
\end{figure}

Since the input to Cyrus is a voice query, we have at least two input data types that we are able to leverage, the sound and the text transcription. Assuming that the voice to text translator was flawless, we can combine analytical tools for both to make sense of the English query in order to map it to an equivalent SQL query. We illustrate the process using the example query \ref{q5} in section \ref{selq}. For this query (superscript shows the word's sequential position in the sentence),
\begin{quote}
{\em Show$^1$ me$^2$ the$^3$ tracks$^4$ composed$^5$ by$^6$ Jimi$^7$ Hendrix$^8$ if$^9$ the$^{10}$ track$^{11}$ id$^{12}$ is$^{13}$ 1479$^{14}$}
\end{quote}
the {\em NSLinguisticTagger} will generate the PoS tags and the universal and enhanced dependencies shown in figure \ref{PosTag}.

To identify the table names, we assign preference to the terms for matching in the enhanced dependency graph and start with the nodes that have the least in-degrees and are not stop words \cite{Stop-Words}, i.e., we consider only {\em Show}$^1$, {\em tracks}$^4$, compounds of {\em Jimi}$^7$ and {\em Hendrix}$^8$ ({\em Jimi Hendrix}), and {\em track}$^{11}$ and {\em id}$^{12}$ ({\em track id}), for which the in-degrees are respectively 0, 1, 2 and 2. We discard {\em Show} because we assume it as an database command and also because it does not match with any database table names ``closely.'' Closeness of a term with another term is determined using a combination of measures such as $\sigma(a,b)= 1- \frac{homo(a,b) + \lambda(a,b) + \psi(a,b)}{3}$, where $homo(a,b)$, $\psi(a,b)$, and $\lambda(a,b)$ respectively are homonym, substring similarity, and Levenshtein similarity functions ranging between $[0,1]$. Function $\psi(a,b)$ is defined as $\frac{sub(a,b)}{|a|}$ such that $sub(a,b)$ returns the length of maximum continuous substring of $a$ of $b$, and $\lambda(a,b)$ is defined as the ratio $\frac{max(|a|,|b|) -lev_{a,b}(|a|,|b|)}{max(|a|,|b|)}$, where higher the value of $\sigma(a,b)$, closer is term $a$ to $b$.

Let us call these terms the set $T_n$ while we call the set of all terms in the sentence without stop words $T_p$. We compute $\sigma(a,b)$ for all terms $a\in T_n$, and for all terms $b\in S$ where $S$ is the set of table names in database $D$, and annotate each term $a$ with their similarity score $\sigma$. We discard all terms $a'\in T_n$ from $T_n$ for which there exists another term $a\in T_n$ such that $\sigma(a,b) \gg \sigma(a',c)$. For example, for query \ref{q5}, we discard the compound term {\em Jimi Hendrix}, but retain {\em track id} along with {\em tracks}, because we have a table in $D$ named {\em Tracks}. We call this new set $T_c$, and call the set of table names in $D$ that matched with the terms in $T_c$, $T_d$. We maintain a list $L_t$ of triples $\langle a, r, \sigma(a,r)\rangle$ such that $a\in T_d$ and $ r \in S$, for every $r$ that met the filter condition $\sigma(a,r) \gg \sigma(a',c)$.

In our next step, for every triple $m\in L_t$ of the form $\langle a, r, \psi\rangle$, we compute a similarity score $\Psi$ as follows.
\[
\Psi(a, r)= \sigma(a,r) + \Sigma_{i=1}^{|T_p|} \mu(a_i,b_i) - \pi(A)
\]
where, $\mu(a_i,b_i)$ is a Stable Marriage \cite{Genc0OS17s} matching function such that pairwise matching is maximized for the terms in $a_i\in T_p$ (i.e., the candidates) and the attributes of the table $r(R)$, i.e., $b_i\in R$. The one to one function $\mu$ assigns $\sigma(a_i,b_i)$ the highest matching score possible such that for no other $b_j\in R$, $\sigma(a_i,b_j) > \sigma(a_i,b_i)$. The terms that could not be matched are assigned 0. If the set of terms $a_i\in T_p$ that could not be matched is $A$, then $\pi(A)=\frac{|A|}{|T_p|-1}$ is a penalty function that compensates for the missing candidates not finding a matching partner. Once we compute $\Psi(a,r)$ for every $m\in L_t$, we choose distinct $r$'s for which $\Psi(a,r)$ is maximum and insert pairs $\langle r, T_p\setminus A\rangle$ in a list $F_t$ of tables names $r$ for the SQL query we intend to construct using attributes $B=T_p\setminus A$, which concludes the process of table and column name identification process.

\subsection{Text to SQL Mapping and Query Processing}

\begin{figure}[h]
	\centering
	\subfigure[Division query \ref{q6}.]{\label{PosTagDiv}\includegraphics[scale=0.335]{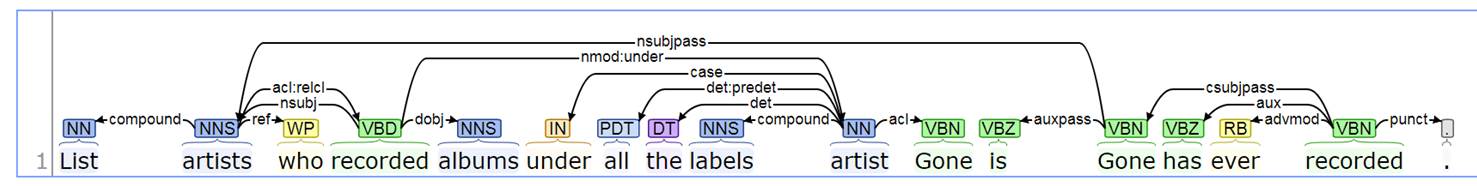}}
	\subfigure[Aggregate/Sum query \ref{q7}.]{\label{PosTagSum}\includegraphics[scale=0.4]{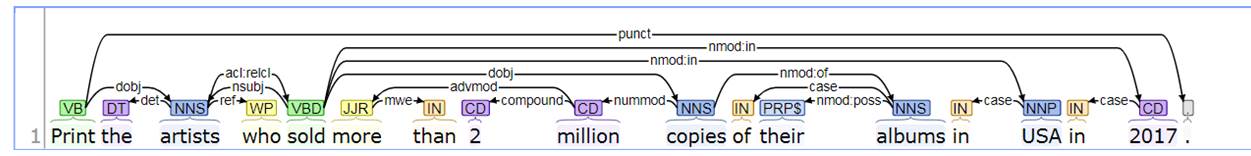}}
	\caption{Dependency parsing.}
	\label{PosTagDiv-Sum}
\end{figure}

In general, construction of an SQL query from an NLQ is a complex and involved task. Especially, when reputed PoS parsers such as Stanford {\em CoreNLP} parser too pose significant challenges in the analysis process. For example, as shown in figure \ref{PosTagDiv}, {\em CoreNLP} parser does not recognize the singer {\em Gone is Gone} as a named entity for query \ref{q6}. Instead, it identifies the artist as three separate verb incarnations. In our heuristic mapping process, we keeping in mind that terms such as ``all'' and ``every'' and their synonyms play important roles in division queries, and that parsers do not always identify all possible named entities, we proceed to analyze the dependency graph cautiously.

Our table and column name identification process will isolate {\em tracks} as the table name, and {\em Artist}, {\em Album}, and {\em Label} as the attributes of interest for table {\em Tracks} during the initial step. For the division query \ref{q6}, we look for a set of objects that are contained in another -- a set of {\em labels.} We do so because the noun ``artists'' record ``albums'', and there is a second term ``recorded'' at the end of $Q_p$, we look to see if other artists are linked. We find in the dependency graph that ``recorded'' is linked to ``Gone'' through a verb ``has'' and a modifier ``ever'', and they are after the second noun ``artist'', we conclude that the set of terms between ``artist'' and ``has'' is a named entity if the {\em NSLinguisticTagger} did not do so already, and we analyze it further. We then discover that the second ``artist'' is qualified -- ``all the labels'', and so we construct the sub-query
\inp
\indent {\sf SELECT} {\em Label}\\
\indent {\sf  FROM} {\em tracks} \\
\indent {\sf WHERE} {\em Artist=} {\em 'Gone is Gone'}
\inp
\noindent The {\em nsubjpass} link from {\em Gone is Gone} back to the first noun ``artist'' helps us construct the container subquery and tie it up with the driver query completing the division query.
\inp
\indent {\sf SELECT} {\em Label}\\
\indent {\sf  FROM} {\em tracks} \\
\indent {\sf WHERE} {\em Artist=} {\em t}
\inp
\noindent Aggregate queries such as query \ref{q7} are constructed similarly by analyzing the dependency graph in figure \ref{PosTagSum}, and realizing that it is also a join query over the attributes {\em Artist, Sale, Region, Album,} and {\em Year} in the tables {\em Tracks} and {\em Charts}. We note that our aggregate query mapping algorithm shares similarity with the approach presented in \cite{ZengLL16s}, without the restrictions they impose on the NLQ.

\section{Conclusion and Future Research}

There are many approaches to advancing NLP-based voice querying. By having an interactive spoken dialogue interface for querying relational databases, we were able to build a functional learning management system for SQL that has the potential to support the classes of queries novice SQL learners encounter frequently. In our first edition of Cyrus, our focus was to build a concept system to demonstrate its feasibility as a voice assistant. In our testing, we have observed that Cyrus was successful in mapping queries \ref{q0} through \ref{q7} properly and executed them flawlessly on our test database. The division and aggregate queries remain under refinement and further adjustment as they are currently minimally functional. Despite its relatively weaker ability to map all division and aggregate queries, the progress Cyrus embodies is nonetheless intellectually satisfying which we plan to address in future. The complete mapping algorithm, which we have omitted in this paper for the want of space, along with a first-database course classroom evaluation remain as our immediate future research.

%\newpage

\bibliographystyle{abbrv}
%\bibliography{/users/jamil/dropbox/bib-db/bib-db-general,/users/jamil/dropbox/bib-db/our-publications}
%\bibliography{/users/hasan/dropbox/bib-db/bib-db-general,/users/hasan/dropbox/bib-db/our-publications}

\end{document}